\documentclass[mathleft
]{an}
\usepackage{graphicx,cite}
\usepackage{times}
\usepackage[figuresright]{rotating}
\usepackage{lscape}

\begin{document}

\def\lesssim{\mathrel{\hbox{\rlap{\hbox{\lower4pt\hbox{$\sim$}}}\hbox{$<$}}}}
\def\gtrsim{\mathrel{\hbox{\rlap{\hbox{\lower4pt\hbox{$\sim$}}}\hbox{$>$}}}}

\Pagespan{789}{}
\Yearpublication{2006}%
\Yearsubmission{2005}%
\Month{11}%
\Volume{999}%
\Issue{88}%

\title{Chandra Imaging of Gamma-Ray Binaries}

\author{O. Kargaltsev\inst{1}\fnmsep\thanks{Corresponding author:
  \email{kargaltsev@gwu.edu}\newline}
  \and B. Rangelov\inst{1}
    \and J. Hare\inst{1}
\and  G.G. Pavlov\inst{2}
}
\titlerunning{Chandra observations of gamma-ray binaries}
\authorrunning{O. Kargaltsev, B.\ Rangelov, J.\ Hare, \& G.G. Pavlov}
\institute{
George Washington University, Department of Physics, 
725 21st St, NW, Washington, DC 20052 
\and 
Pennsylvania State University, Department of Astronomy and Astrophysics,
University Park, PA 16802
}

\received{24 Aug 2013}
\accepted{11 Nov 2013}
\publonline{later}


\abstract{We review the multiwavelength properties of  the few known gamma-ray binaries, focusing on extended emission recently resolved with \emph{Chandra}. We discuss the implications of these findings for the nature of compact objects and for physical processes operating in these systems.}

\maketitle

\section{Introduction}

Thanks to  recent advances in space-based X-ray, high energy (HE, GeV), and very high energy (VHE, TeV) $\gamma$-ray observations, an emerging population of high-mass gamma-ray binaries (HMGBs) has become an important topic in modern high-energy astrophysics. The five VHE HMGBs firmly detected in TeV and/or  GeV $\gamma$-rays, can be observationally separated into two types: (1) binaries where a rotation-powered pulsar interacts with the strong wind of the massive stellar companion (LS~2883/B1259--63), and (2) microquasars (LS~5039 and LS~I~+61 303). The types of the other two HMGB, HESS~J0632+057 and HESS~J1018--589A/1FGL~J1018.6-5856, are so far uncertain. The latter is  the only HMGB  coincident with the supernova remnant (G284.3--1.8). In addition to the five VHE  HMGBs  there are several other massive binary systems (e.g., Cygnus~X-3,  $\eta$~Carina) which  were detected in the GeV band with {\sl Fermi} LAT but  so far lack the TeV detections. In most of the VHE binaries the GeV light curves are modulated with the orbital period determined from optical spectra and/or radio measurements. Below we summarize and compare the observational properties of the HMGBs focusing on imaging observations of VHE HGMBs with the {\sl Chandra X-ray Observatory (CXO)}. We also discuss possible reasons for different multiwavelength (MW) manifestations of these systems. 
 
 \section{MW Properties of  HMGBs}
 
\begin{figure}
\centering{\includegraphics[width=0.92\hsize,trim=15 0 15 0]{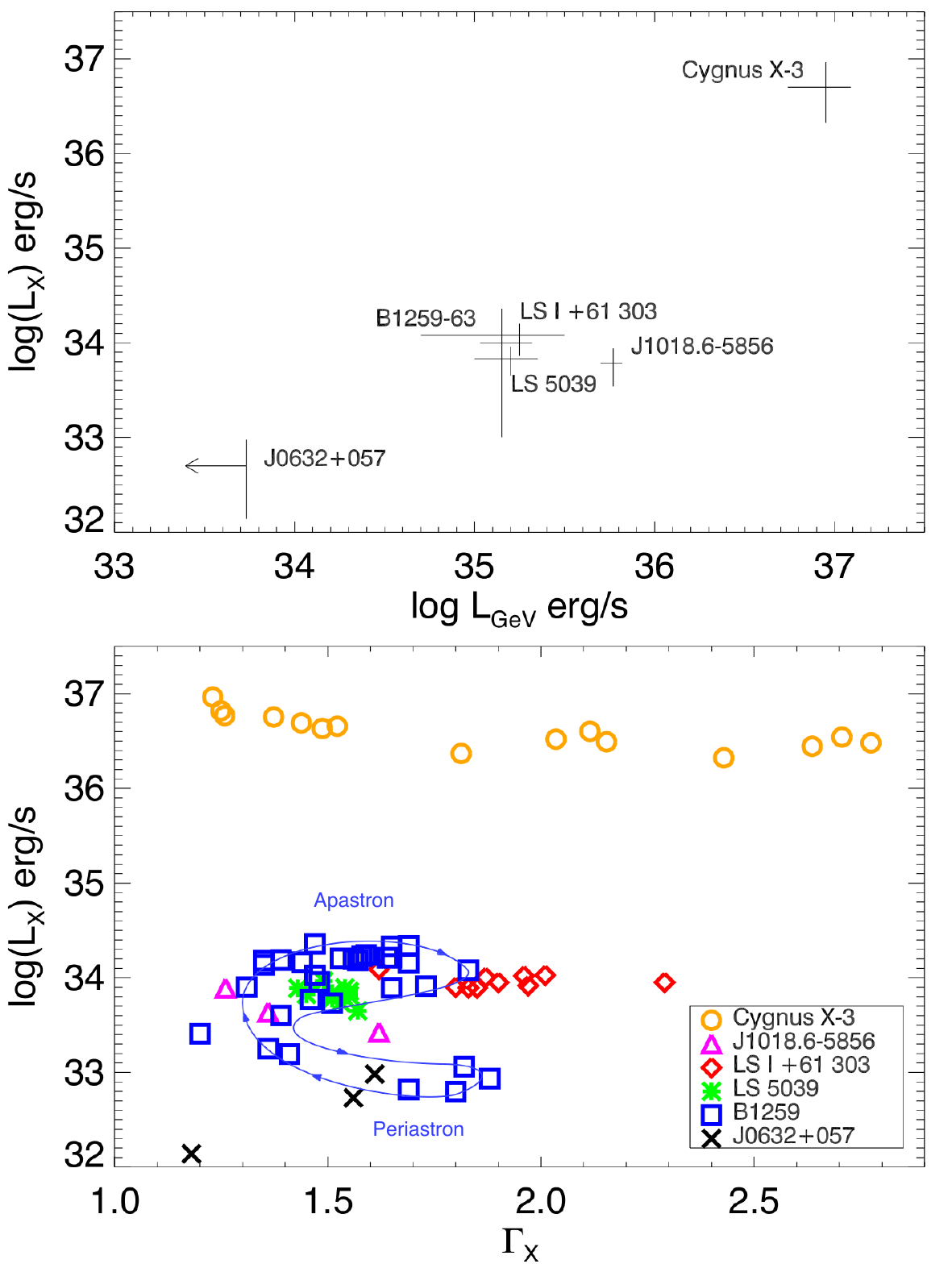}}
\caption{X-ray luminosity $L_X$ vs. GeV luminosity $L_{GeV}$ (\emph{top}), and $L_X$ vs. $\Gamma_X$ (\emph{bottom}). The lines represent the observed parameter ranges as listed in Table~1. The arrow on J0632+057 (\emph{top} panel) shows the upper limit of its GeV luminosity. For Cygnus~X-3 we only show a subset of the available data (\cite{2010ApJ...718..488S}) for which satisfactory fits were obtained (excluding extremely hard or soft components). The blue curve with arrows (\emph{bottom}) shows the evolution of B1259 parameters with the orbital phase.}
\label{MW}
\end{figure}

\begin{figure*}
\centering{\includegraphics[width=0.507\hsize]{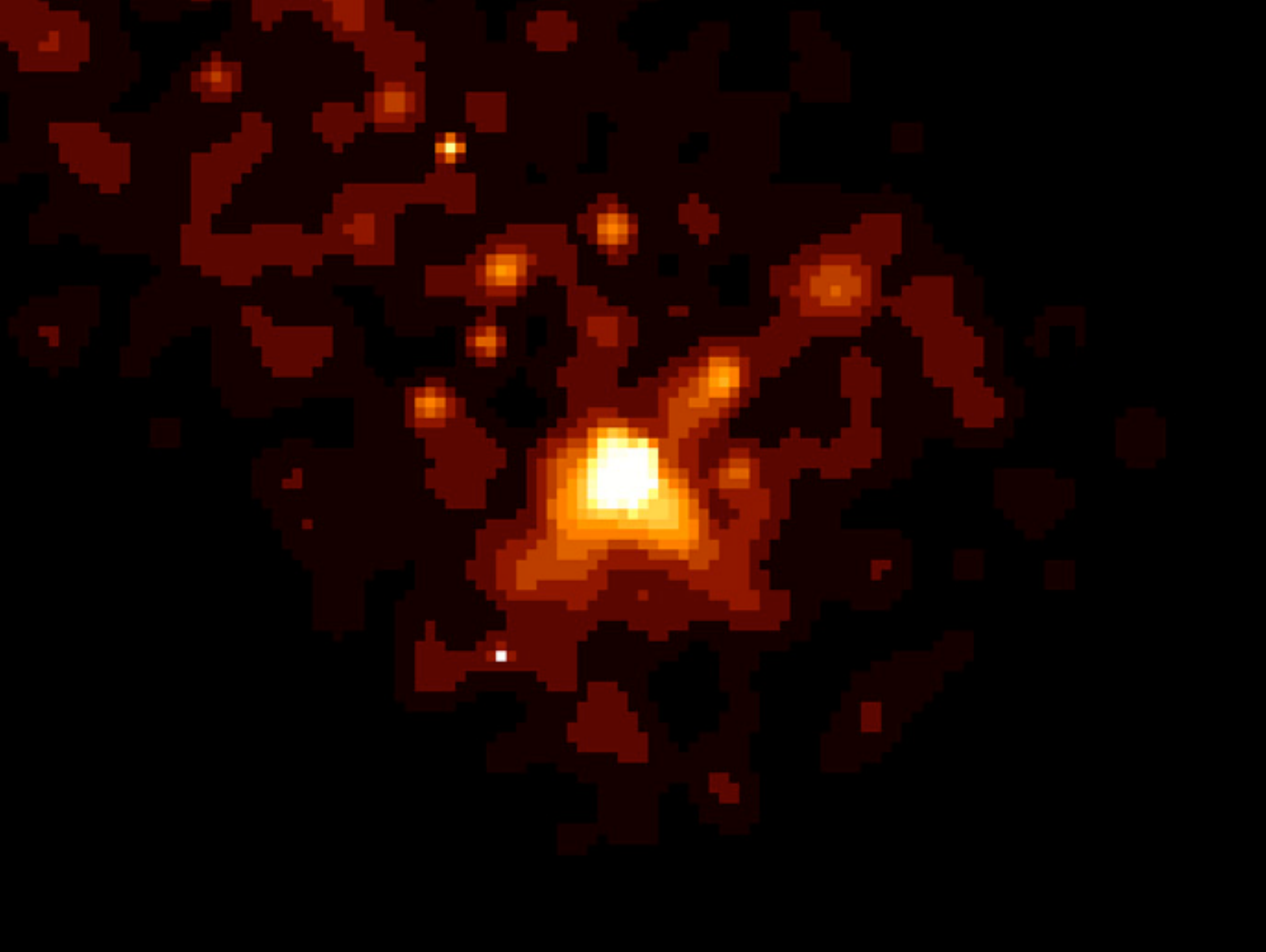}
\includegraphics[width=0.485\hsize]{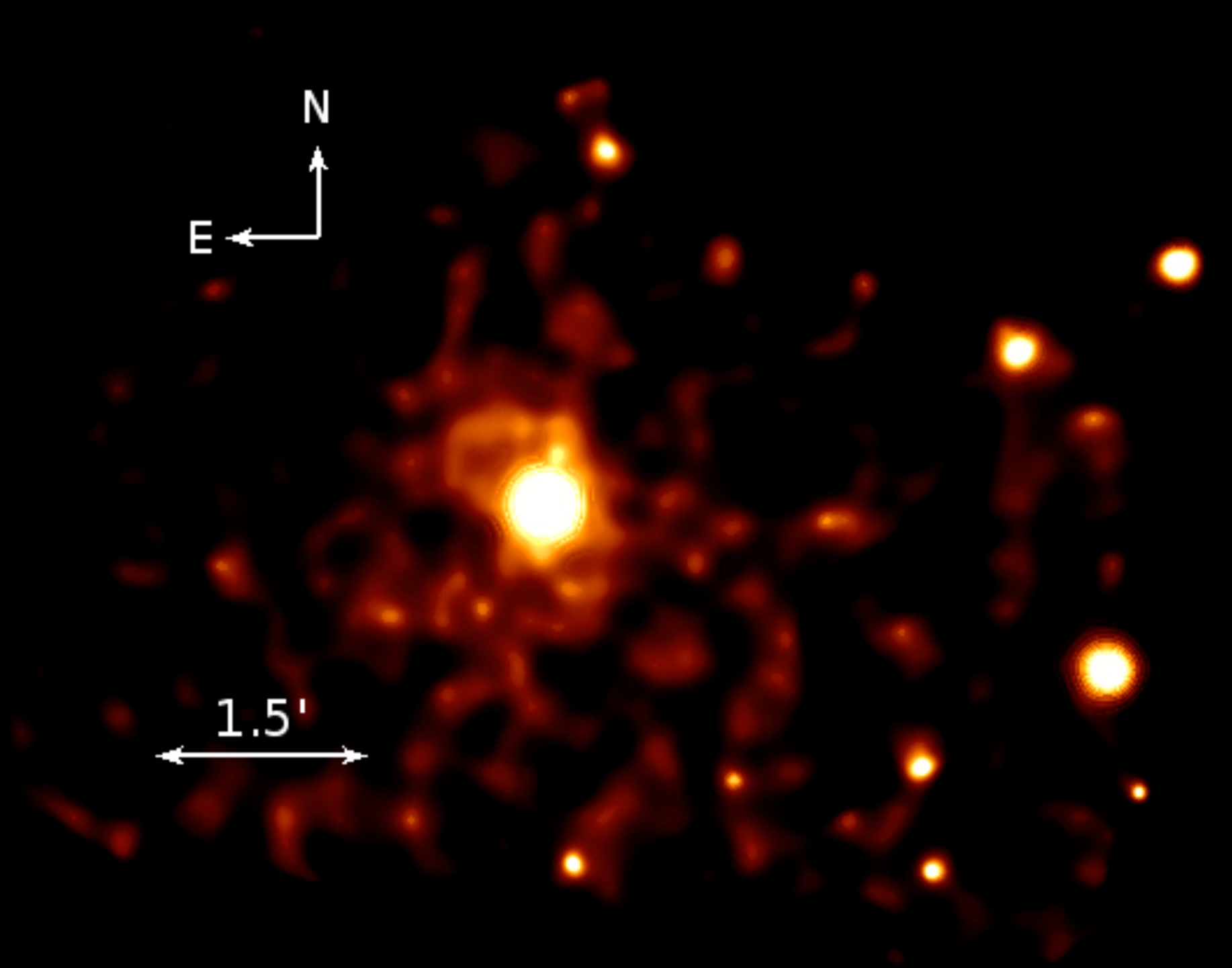}}
\caption{\emph{Left:} 28~ks ACIS-I3 image ($0.5-8$~keV; binned to a pixel size of $1''$ and adaptively smoothed) showing  extended emission surrounding LS~2883 (adopted from \cite{2011ApJ...730....2P}). \emph{Right:} 38~ks ACIS-I image ($0.7-7$~keV; binned to a pixel size of $1''$ and adaptively smoothed) of LS 5039 vicinity  (see also \cite{2011ApJ...735...58D}).
\label{LS5032}}
\end{figure*}
 
All five VHE HMGBs are fairly bright in GeV $\gamma$-rays  and X-rays with the exception of HESS~J0632+057 whose GeV emission (\cite{Hill2013}) has not been detected yet (the limit on its GeV luminosity is lower than the GeV luminosity of the other 4 VHE HMGBs). However, given the somewhat lower TeV and X-ray luminosities of HESS~J0632+057, it is possible that the distance to this HMGB ($d=1.5$ kpc; \cite{2007A&A...469L...1A}) is underestimated. 

We compare the MW properties of the HMGBs presented in Table~1. While the small number of   HMGBs prevents us from drawing any solid conclusions about this population of $\gamma$-ray emitters, we still notice some correlations between their MW properties (see also \cite{2013arXiv1307.7083D} for more detailed discussion of MW properties). Fig.~\ref{MW} shows the X-ray luminosity as a function of the GeV luminosity (top) and $\Gamma_X$ (bottom). Although the sample is small,  one  can notice from Fig.~1 that B1259, LS~5039, and LS~I~+61~303 inhabit the same locality on the plot, which suggests that they may be similar in their nature. We can also note that their $L_{\rm TeV}$  and $\Gamma_{\rm TeV}$ are similar to those of TeV PWNe (see Fig.~7 from \cite{2013arXiv1305.2552K}). This would not be surprising if the source of power in these systems were a young pulsar, and the $\gamma$-ray radiation were due to ICS of the starlight. Unfortunately, with the exception of  LS~2883/B1259--63, the nature of compact objects in the other four $\gamma$-ray binaries has not been confidently established yet. 

The {\sl Fermi} LAT spectra of the two microquasars, LS~5039 and LS~I+61~303, and 1FGL~J1018.6--5856 exhibit  cut-offs in 2--6 GeV range when modeled as power-laws ($\Gamma=1.9-2.2$) with exponential cut-offs.  This  is very similar to the GeV spectra of isolated pulsars. These systems show only modest variability levels in  GeV compared to LS~2883 (\cite{2012ApJ...749...54H}). The more variable LS~2883, which is known to host the young pulsar B1259--63, shows a noticeably harder spectrum ($\Gamma\approx1.4$) and substantially lower cut-off energy  ($\approx0.3$ GeV)  in the bright flaring state (\cite{2011ApJ...736L..11A}). Cygnus X-3 appears to exhibit a much softer ($\Gamma=2.5-2.7$) spectrum when it is detected through its flares  in GeV (\cite{2012MNRAS.421.2947C}).

\section{Extended X-ray emission.}

The exquisite angular resolution of {\sl CXO} resolved extended emission from HMGBs for the first time (Fig.~\ref{LS5032}). We consider these findings secure for two HMGBs (LS~2883, LS~5039) while in  other cases additional observations are needed to  firmly establish  the existence and the origin of the extended emission (see below). Since HESS~J0632+057 has not been observed in the imaging mode with {\sl CXO} few constraints exist regarding the presence of extended emission. Based on  similarities between  HESS~J0632+057 and LS~2883 (i.e., long period, large eccentricity, and hints of extended radio emission), one may expect to detect such emission around  HESS~J0632+057 as well, if the compact object is also a pulsar.  The existing {\sl CXO}  observations of all HMGBs are summarized in Table~2. Radio observations of HMGBs also revealed extended (and sometimes variable) structures (\cite{2013arXiv1306.2830M})  which however have much smaller scales ($\sim$3--500 AU)  than the X-ray nebulae described below.  
 
\noindent  {\bf LS~2883/B1259--63:} Extended emission around LS~2883/B1259--63  has been reported by \cite{2011ApJ...730....2P}  from {\bf a} 28~ks {\sl CXO} observation (see Fig.~\ref{LS5032} left). It has been suggested that the emission could be a shocked pulsar wind escaping the binary near apastron. Since then two more {\sl CXO} observations have been carried out. The extended emission was detected with high confidence and was found to be variable both in morphology and flux. The detailed results  of these two observations are discussed in our forthcoming paper. 
 
\noindent {\bf LS~5039:} \cite{2011ApJ...735...58D} analyzed the 38~ks \emph{CXO} ACIS observation and found evidence of nonthermal emission extending up to $1'$  from LS 5039 (see Fig.~\ref{LS5032}, right). The spectrum could be fitted by an absorbed power-law (PL) model with $\Gamma = 1.9\pm0.3$  and unabsorbed 0.5--8 keV luminosity $L_{X}\simeq1\times10^{32}$~erg~s$^{-1}$ for a plausible distance of $2.5$~kpc. 

\begin{landscape}
\begin{table}
\setlength{\tabcolsep}{0.045in}
\caption{Multiwavelength properties of HMGBs with compact objects. $^a$ Galactic longitude. $^b$ Galactic latitude.$^c$ Compact object. $^d$ Companion Star. $^e$ Binary period. $^f$ Binary orbit eccentricity. $^g$ Distance. $^h$ Logarithm of luminosity in $1-10$~TeV band (range for variable sources). $^i$ Photon index of TeV spectrum. $^j$ Logarithm of luminosity in the $0.1-1$~GeV band (range for variable sources). $^k$ Photon index of GeV spectrum.$^l$ Logarithm of X-ray luminosity in $1-10$~keV band (range for variable sources).$^m$ Photon index of X-ray spectrum. $^n$ Extended emission in Radio/X-rays (Y=YES, N =NO, P=Possibly, ?= Unknown). $\newline$ $^\star$ Measured in the flaring phase rather than throughout the entire orbit.  $^o$ References: 1. \cite{2007A&A...469L...1A,2011ApJ...737L..12R,Hill2013}; 2. \cite{2013A&A...551A..94H, 2011ApJ...730....2P,2011ApJ...736L..11A,2009MNRAS.397.2123C}; 3. \cite{2005Sci...309..746A,2009ApJ...697L...1K, 2012IAUS..282..331S,2012ApJ...749...54H,2009ApJ...698..514A}; 4. \cite{2008ApJ...679.1427A,2009ApJ...693.1621S,2012ApJ...749...54H,2009ApJ...698..514A}; 5. \cite{2012A&A...541A...5H,2012Sci...335..189F}; 6. \cite{2009arXiv0908.0714G,2013arXiv1307.3264B,2008AIPC.1054...13H,2009Sci...326.1512F}}
\begin{center}
{
\begin{tabular}{clccccccccccccccc}

\hline

\# & Source & $l^a$ & $b^b$ & CO$^c$ & Comp.$^d$ & $P_{\rm orb}^e$ & $e^f$ & $d^g$ & $\log L_{\rm TeV}$$^h$ & $\Gamma_{\rm TeV}$$^i$ & $\log L_{\rm GeV}$$^j$ & $\Gamma_{\rm GeV}^k$ & $\log L_{\rm X}$$^l$ & $\Gamma_{\rm X}$$^m$ & Ext.$^n$ & Ref.$^{o}$ \\

& & deg& deg& & & d & & kpc& erg/s & & erg/s & & erg/s  \\

\hline

1. & HESS~J0632+057 & 205.66 & $-1.44$ & NS? & B0pe & $321$ & $0.8$ & 1.5 & 32.7 & 2.5 & $\lesssim 33.7$ & ... & 32.14--32.98 & 1.2--1.6 & Y/? & 1\\

2. & LS~2883/B1259--63 & 304.19 & $-0.99$ & PSR & O8V & 1236 & 0.87 & 2.3 & 32.96$^{\star}$ & 2.9 & 34.70$^{\star}$--35.50$^{\star}$ & 1.4--2.4 & 33.0--34.36 & 1.2--1.9 & Y/Y & 2\\

3. & LS~5039 & 16.90 & $-1.28$ & ? & O7V & 3.9 & $0.34$ & 2.5 & 34.09 & 2.12 & 35.00--35.35 & 2.0--2.07 & 33.65--33.96 & 1.45--1.57& Y/P & 3\\

4. & LS~I~+61~303 & 135.67 & 1.09 & NS? & B0Ve & 26.5 & 0.54 & 2.0 & 32.93 & 2.4 & 35.03--35.32 & 2.0--2.2& 33.86--34.20 & 1.25--2.1 & Y/P & 4\\

5. & 1FGL~J1018.6--5856 & 284.26 & $-1.82$ & ? & O6V & 16.6 & ? & 5.0 & 33.36 & 2.70 & 35.70--35.82 & 1.9 & 33-54--33.94 & 1.26--1.36 & ?/P & 5\\

6. & Cygnus~X-3 & 292.09 & 0.34 & BH? & WR & 0.2 & 0 & 9 & $<33.8$ & ... & 36.74--37.09 & 2.5--2.7 & 36.4-37.2 & ... & Y/N & 6

\end{tabular}}
\end{center}

\label{tab:prop4}

\end{table}

\begin{table}
\setlength{\tabcolsep}{0.06in} 
\caption{Summary of existing \emph{CXO} observations and extended X-ray emission properties.}
\begin{center}
{
\begin{tabular}{clcccccccccccc}
\hline
 & Binary name & Detector:Mode/Exposure (in ks)&  Extended?  & Extent& Luminosity  & Photon index& References  \\
\hline
 &  &   &   &  & 10$^{31}$ erg s$^{-1}$ & \\
\hline
1 & HESS~J0632+057 &  ACIS:CC/40 &  ? & ... & ... & ... & 1 \\
2 & LS~2883/B1259--63 & ACIS:TE/138,HETG/9,CC/10 &   Y   & $\sim6''$ & 2--6 & 1.3--1.5 &  2  \\
3 & LS~5039  &   ACIS:TE/38,HETG/11,CC/70 &   Y   & $\sim1'$ & 6.6 & $\sim1.9$ & 3 \\   
4 & LS~I~+61~303  &  ACIS:TE/72,CC/95 &   P   & $\sim12''$ & $\sim2$ & ... &  4 \\
5 & 1FGL~J1018.6--5856  &  ACIS:TE/10 &   P   & $\sim 4'$ & 20--90& ... & ... \\
6 & Cygnus~X-3  &   HRC:I/15; ACIS:HETG/177 &   N  & ... & ... & ... &  ... 
\end{tabular}}
\end{center}
\label{tab:prop4}
\textbf{References:} 1. \cite{2011ApJ...737L..12R}; 2. \cite{ 2009MNRAS.397.2123C,2011ApJ...730....2P}; 3.  \cite{ 2011ApJ...735...58D,2011MNRAS.416.1514R,2005A&A...430..245M}; 4. \cite{2007ApJ...664L..39P,2010MNRAS.405.2206R}
\end{table}
\end{landscape}

\noindent The non-isotropic morphology and the hard spectrum of the  observed emission argue against the dust halo (which was also detected in this observation on a smaller angular scale). The presence of the nonthermal nebula suggests that the compact object is a pulsar. If so, the termination shock in the pulsar wind would probably occur at much smaller distances than those typical for isolated pulsars with PWNe, and, since the binary orbit is rather tight, the pulsar wind should  interact with the denser  O-star wind (compared to LS~2883). Such interaction and the accompanying  mixing due to the instabilities would probably prevent the formation of any fast bulk flow (\cite{2011A&A...535A..20B}), however, some ultra-relativistic electrons can still  escape via diffusion to large distances before their energies become too low to emit synchrotron X-rays.

\begin{figure*}
\centering
\vbox{
\parbox{0.5\hsize}{
\includegraphics[width=\hsize]{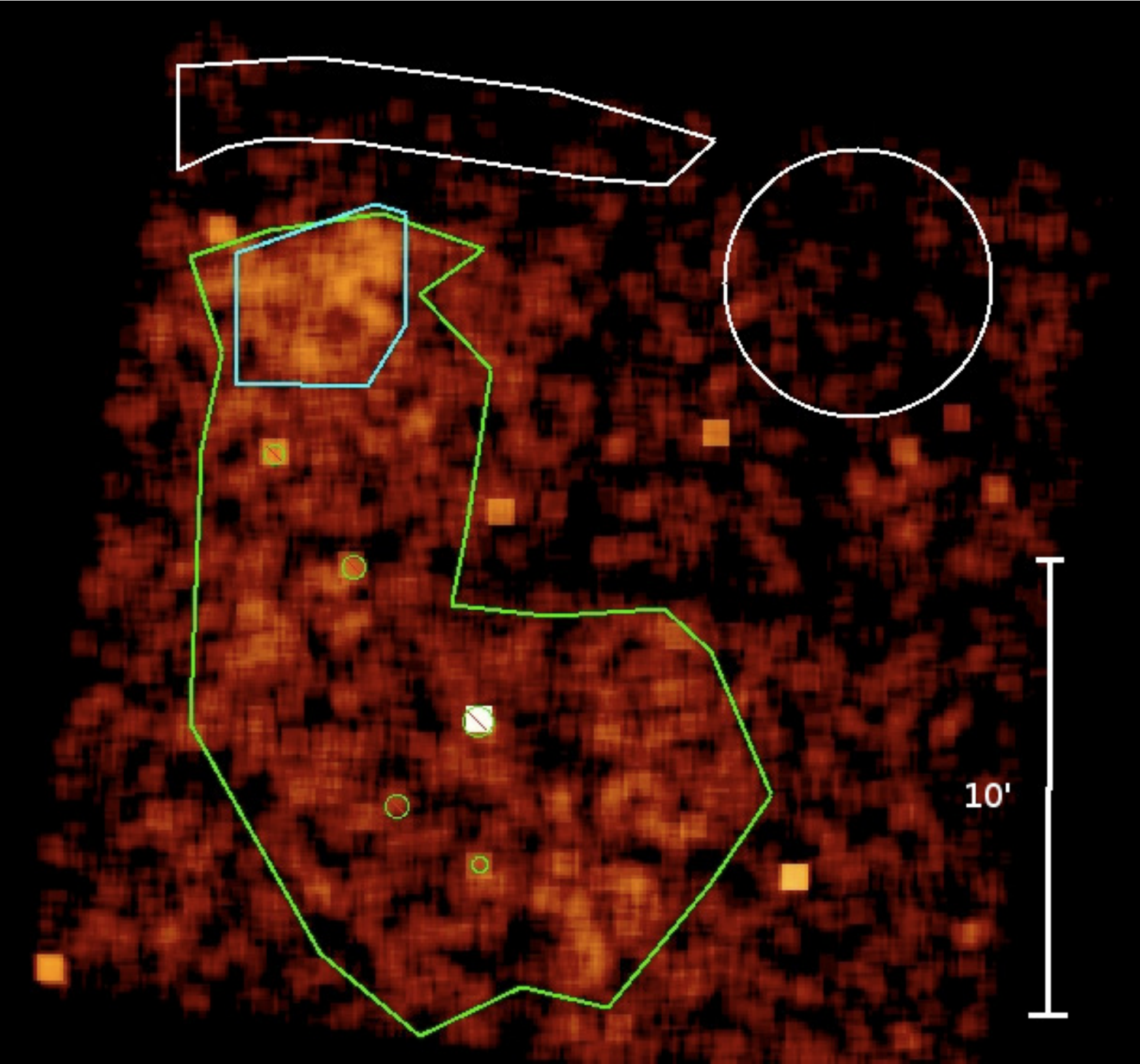}}
\parbox{0.45\hsize}{
\includegraphics[width=\hsize]{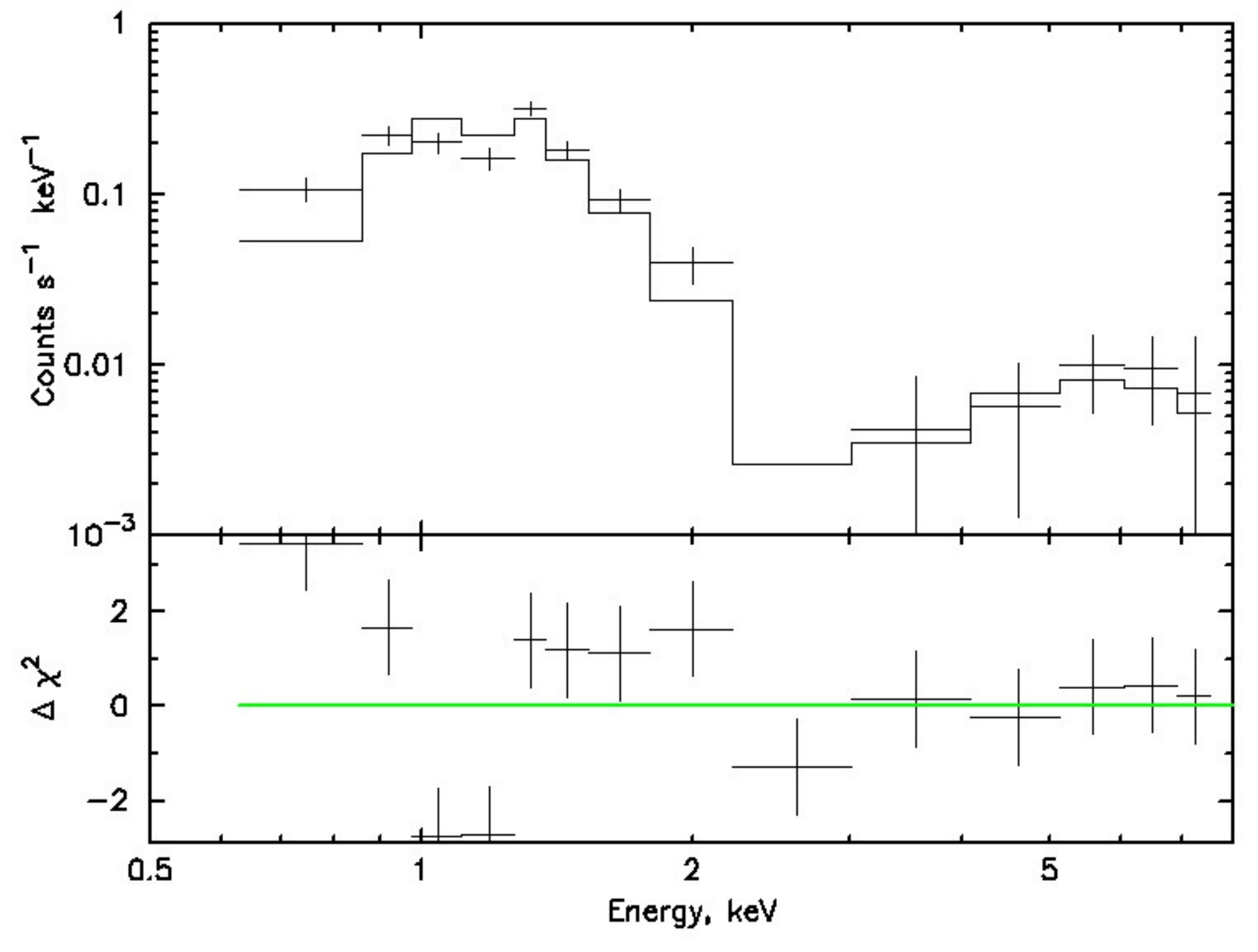}
\includegraphics[width=\hsize]{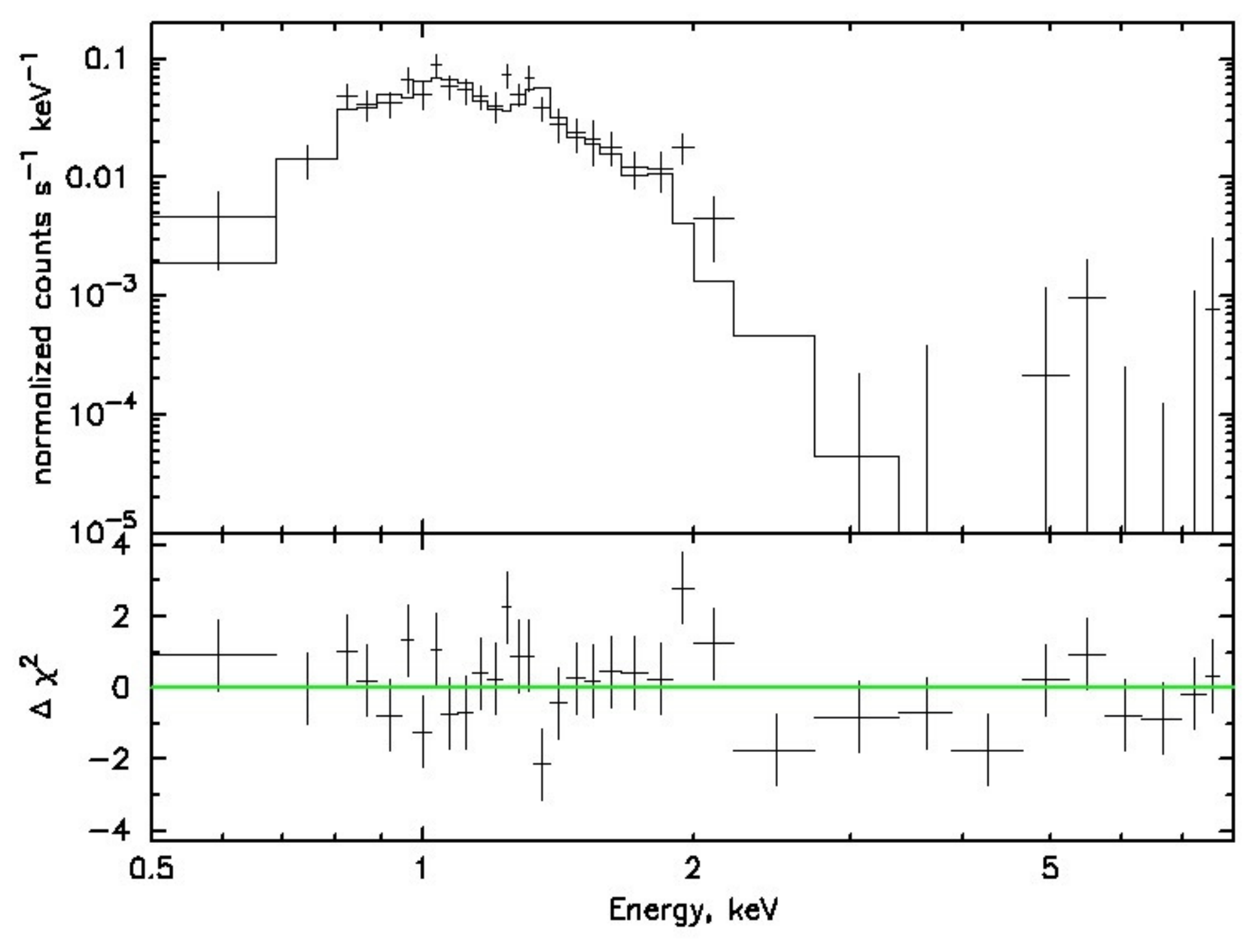}}
}
\caption{\footnotesize
{\em Left:} ACIS-I 10\,ks image (0.5--8\,keV) of the field of 1FGL~J1018--5856. Regions for preliminary spectral analysis of the apparent extended emission are shown: the green line encompasses the entire diffuse emission, and the blue line only the soft northern lobe. Appropriate background regions are marked in white. Point sources were excluded from the extraction region, e.g., J1018 (the brightest source near the centre).
{\em Right:} Spectral fit to the whole nebula (MEKAL+PL model, top) and the soft northern lobe (MEKAL, bottom). The MEKAL temperatures are similar for both regions (0.22\,keV and 0.23\,keV, respectively), but only the whole nebula spectrum requires a hard component. Although significant, its photon index is very poorly defined.
}
\vspace{-0.5cm}
\label{field}
\end{figure*}

\begin{figure*}
\centering
\hspace{-0.7cm}
\vbox{
\includegraphics[width=0.98\hsize]{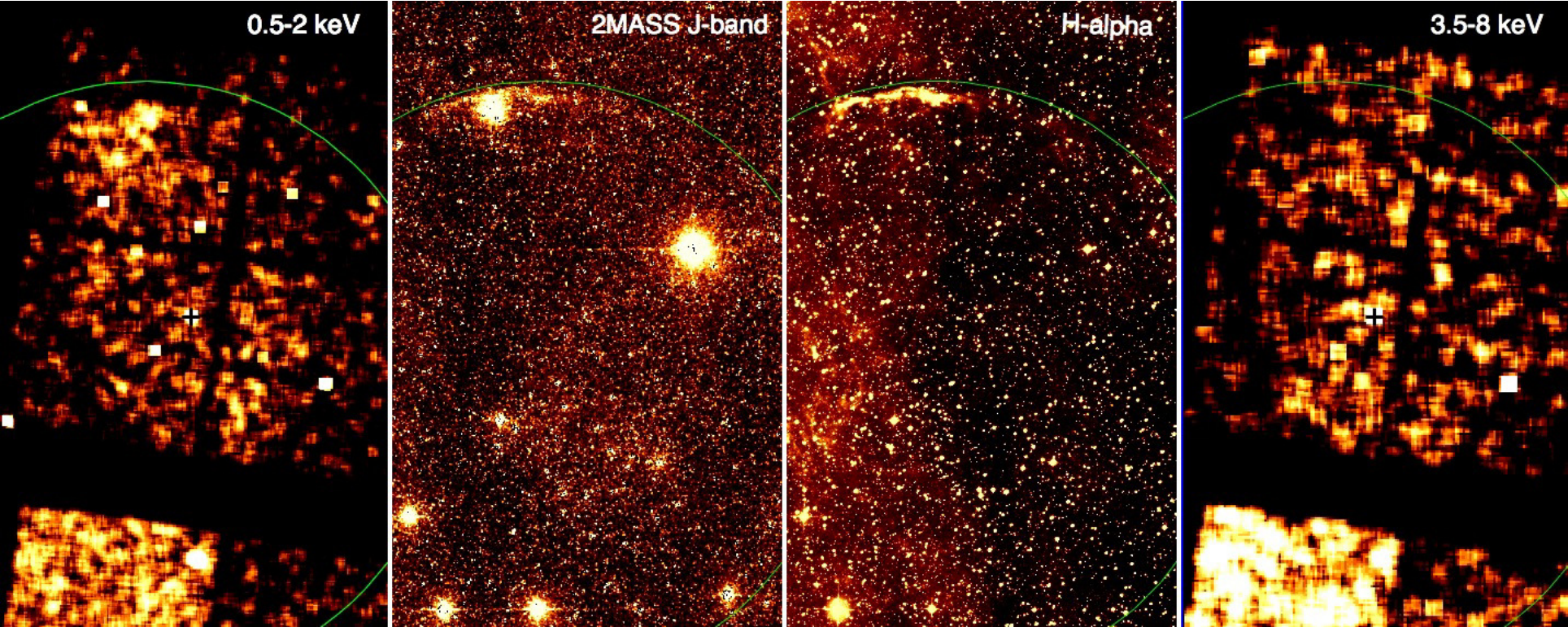}
}
\caption{\footnotesize
Field in the vicinity of J1018 in two X-ray bands, with auxiliary NIR-optical data. Of particular interest is the filament seen in the two central panels, which appears to be a section of a circle and located just beyond the extent of the X-ray emission.}
\label{aux}
\end{figure*}

\noindent {\bf 1FGL~J1018.6--5856:} The donor star  in this binary is similar to that of LS 5039, but the orbital period is a factor of 4 longer, and the distance is likely a factor of two larger. Similar to  LS~5039, arcminute-scale extended emission is seen around the binary (see Fig.~\ref{field}) in the 10~ks {\sl CXO} observation (\cite{2011ATel.3228....1P}). The observed spectrum  consists of two components, the softer of which is likely the SNR emission. The shell of G284.3--1.8 is clearly seen to the north of the binary (see Fig.~\ref{aux}), and the X-ray emission adjacent to the shell in the north  has a soft X-ray spectrum compatible with the softer component of the spectrum extracted from the vicinity of 1FGL~J1018.6--5856. The origin of the harder X-ray component is so far unclear. Compared to LS 5039, the angular extent of the hard emission is a factor of 4--5 larger (at a factor of 2 larger distance) while the spectrum appears to be much harder than that of the extended emission around  LS~5039 (see Fig.~\ref{field}).   The forthcoming 80 ks {\sl CXO} observation will shed more light on this interesting source.

\noindent  {\bf  LS~I~+61~303:} The extended emission has been reported by \cite{2007ApJ...664L..39P} based on a 50~ks {\sl CXO} observation carried out with the ACIS-I array (\cite{2007ApJ...664L..39P}). It was interpreted as possibly being thermal Bremsstrahlung created from a large-scale outflow from LS I +61 303 which interacts with the nearby  hot and dense ISM. Another possible explanation is that this is non-thermal synchrotron or inverse Compton emission produced by particles beyond the termination region of the pulsar wind (\cite{2007ApJ...664L..39P}).The presence of the extended emission has been disputed by \cite{2010MNRAS.405.2206R} who argued that there was no significant extension observed from this source based on the ACIS continuous clocking (CC) mode  data, which, however, only offers one-dimensional imaging capability and suffers from a much larger background compared to the conventional imaging.

\noindent {\bf Cygnus~X-3:} 
 Cyg~X-3 does not show extended emission in the {\sl CXO} images which indicates a lack (or reduced numbers) of ultra-relativistic electrons. Cyg~X-3 shows orbital modulation in both soft and hard states that is stronger in softer states. The modulation amplitude decreases with increasing energy in the hard, intermediate, and very high states (\cite{2013ApJ...763...34W}). The sporadic GeV emission  reported by \cite{2013arXiv1307.3264B}, and  \cite{2012MNRAS.421.2947C}, is likely to be associated with the state transitions although some  orbital phases has  also been seen (\cite{2009Sci...326.1512F}). The existence of extended radio jets, which qualifies Cyg~X-3 as a microquasar, is common for both BH and NS X-ray binaries (the former is favored for Cyg~X-3 when both radio and X-ray emission are considered (\cite{2010MNRAS.402..767Z}).  However, the presence of the jets (the defining feature of  microquasars), often associated with the particle acceleration sites, does not appear to guarantee copious GeV or TeV $\gamma$-ray production neither for BH nor for NS HMXBs.  
 
Finally, the most populous group of HMXBs which are not microquasars are the accreting X-ray pulsars (such as the wind-fed Vela X-1 or 1A0535+262) which do not show nonthermal components extended to $\gamma$-rays.  X-ray emission from these systems is dominated by thermal emission with Comptonized hard X-ray tail and  powered by either accretion disk or hot accreting column above the surface.  The apparent deficit of ultra-relativistic electrons in these systems suggests that they are inefficient accelerators. 

\acknowledgements
This work was partly supported by Chandra award GO2-13085 and by NASA grant NNX09AC81G.


\end{document}